\documentclass[journal=esthag,manuscript=communication]{achemso}
\usepackage[version=3]{mhchem} 
\usepackage{graphicx}

\title[An \textsf{achemso} demo]
  {Water Purification by Shock Electrodialysis: Deionization, Filtration, Separation, and Disinfection}
\author{Daosheng Deng}
\affiliation{Department of Chemical Engineering, Massachusetts Institute of Technology, Cambridge, MA, USA}
\author{Wassim Aouad}
\altaffiliation{Present address: St Edmund's College, University of Cambridge, Cambridge, UK}
\author{William A. Braff}
\affiliation{Department of Mechanical Engineering, Massachusetts Institute of Technology, Cambridge, MA, USA}
\author{Sven Schlumpberger}
\affiliation{Department of Chemical Engineering, Massachusetts Institute of Technology, Cambridge, MA, USA}
\author{Matthew E. Suss}
\affiliation{Department of Chemical Engineering, Massachusetts Institute of Technology, Cambridge, MA, USA}
\author{Martin Z. Bazant}
\affiliation{Department of Chemical Engineering, Massachusetts Institute of Technology, Cambridge, MA, USA}
\alsoaffiliation{Department of Mathematics, Massachusetts Institute of Technology, Cambridge, MA, USA}
\email{bazant@mit.edu}

\begin{document}
\clearpage
\begin{abstract}
The development of energy and infrastructure efficient water purification systems are among the most critical engineering challenges facing our society. Water purification is often a multi-step process involving filtration, desalination, and disinfection of a feedstream. Shock electrodialysis (shock ED) is a newly developed technique for water desalination, leveraging the formation of ion concentration polarization (ICP) zones and deionization shock waves in microscale pores near to an ion selective element. While shock ED has been demonstrated as an effective water desalination tool, we here present evidence of other simultaneous functionalities. We show that, unlike electrodialysis, shock ED can thoroughly filter micron-scale particles and aggregates of nanoparticles present in the feedwater. We also demonstrate that shock ED can enable disinfection of feedwaters, as approximately $99\%$ of viable bacteria (here \textit{E. coli}) in the inflow were killed or removed by our prototype. Shock ED also separates positive from negative particles, contrary to claims that ICP acts as a virtual barrier for all charged particles. By combining these functionalities (filtration, separation and disinfection) with deionization, shock ED has the potential to enable more compact and efficient water purification systems. 
\end{abstract}
\clearpage
\section{INTRODUCTION}
The purification of sea or brackish water is an increasingly important process in areas suffering from water stress or scarcity \cite{shannon2008}. State of the art water purification is performed primarily by reverse osmosis (RO) plants and in some cases by electrodialysis (ED) plants \cite{shannon2008, chao2008feasibility}. In RO and ED plants the complete water purification process can be roughly divided into three sequential steps: i) upstream feedwater processing, ii) salt removal (desalination), and iii) downstream processing of the product water \cite{sauvet2007ashkelon, elimelech2011future, semiat2008energy}. In RO plants, to perform desalination, the feedwater is pressurized to above its osmotic pressure and flows through an RO membrane which inhibits the transport of salts. In ED, feedwater flows through an open channel between an anion and cation exchange membrane, and applying an appropriately directed ionic current to this system causes anions and cations (dissolved salts) to exit the space between the membranes~\cite{probstein1994,nikonenko2010}. Many of the upstream steps in the water purification process are to prevent fouling and scaling of the RO or ED membranes, and these include filtration to remove slit (membrane foulants), and pH adjustments of the feedwater \cite{greenlee2009reverse}. Downstream processes include disinfection of the desalted water through use of chemical additives such as chlorine \cite{greenlee2009reverse}. Modern RO plants typically require roughly 4 kWh/m$^3$ of freshwater to complete the purification process from intake sea water to potable water \cite{semiat2008energy,sauvet2007ashkelon}, with the upstream and downstream processes requiring together roughly one third of this total energy.\cite{elimelech2011future,semiat2008energy}

\begin{figure}
\includegraphics[width=5in]{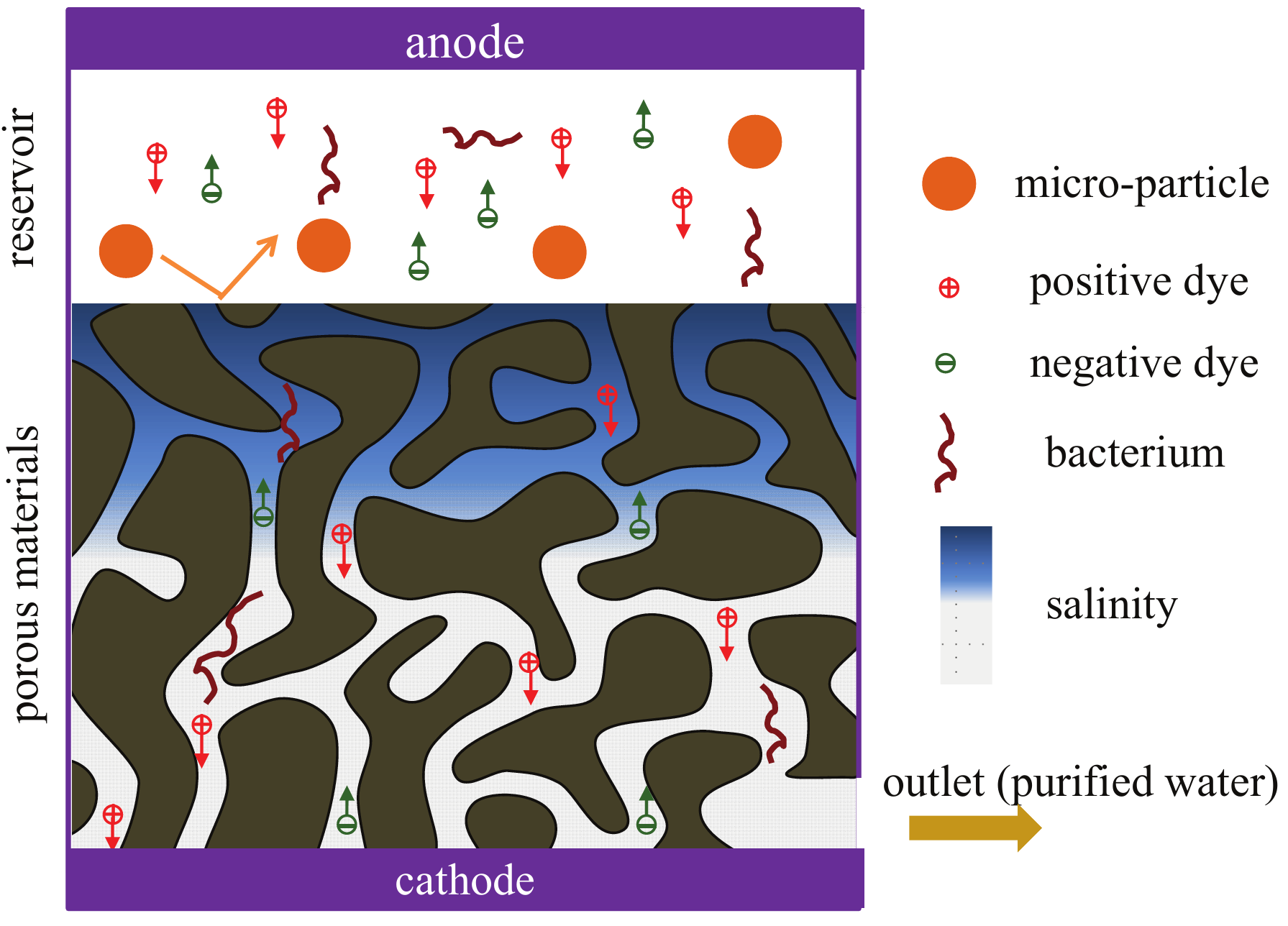}
\caption{Schematic demonstrating water purification with our shock ED device. Our shock ED cell consists of two ion selective elements (electrodes or ion exchange membranes) between which is placed a porous media. By passing an ionic current between the ion selective elements, a salt depletion zone is formed near to the cathode. In addition to leveraging the depletion zone to produce desalinated water, the device demonstrates other unique functionalities, including filtration of particulates, spatial separation of species by valence sign, and disinfection.}
\label{fig:sketch}
\end{figure}

Shock electrodialysis (shock ED) is a promising new technique for water desalination which differs from ED in several key aspects.\cite{deng2013} In its current realization, a shock ED cell consists of two ion selective elements (ion exchange membranes or electrodes) between which feedwater flows through a charged porous medium with thin double layers that acts as a ``leaky membrane"~\cite{dydek2013,yaroshchuk2012acis} (Figure~\ref{fig:sketch}).\cite{deng2013} Like ED, when current is passed through the shock ED cell, an ion depleted zone is formed along an ion selective element (the cathode in Figure~\ref{fig:sketch}). As current is increased, ion concentration near this element approaches zero, and the system can reach the classical diffusion-limited current.\cite{probstein1994} However, unlike ED, in shock ED the presence of a surface charge along the porous media's internal surfaces can enable transport of ions faster than diffusion, an ``over-limiting current", via surface conduction and electro-osmotic flow.\cite{dydek2011} As a result, the depletion zone can be propagated through the pores as a shock (i.e. with a sharp boundary between the depleted and undepleted zones).\cite{mani2009,zangle2009,mani2011,suss2011,yaroshchuk2012acis,dydek2013} Water flowing through the depletion zone emerges from the cell as desalinated water.\cite{deng2013} In shock ED, an ion enrichment or brine zone is formed at the opposite ion selective element, and the formation of enriched and depleted zones at opposite ends of the cell is termed (ion) "concentration polarization" (ICP).\cite{probstein1994,nikonenko2010} 

Previously, we developed a shock ED prototype using a porous silica glass frit with micron-scale pores as the porous medium, a copper electrode as the anode-side ion selective element, and a Nafion ion exchange membrane as the cathode-side ion selective element.\cite{deng2013} With this device, we demonstrated the deionization of a copper sulfate solution by reducing its concentration by roughly 4 orders of magnitude in two passes (to 10 $\mu$M). Further, our measurements of overlimiting conductance suggested that the overlimiting current mechanism in our prototype device was electroosmotic flow rather than surface transport. Compared to recently-developed microfluidic approaches leveraging ICP for water desalination,\cite{kim2010,knust2013} shock electrodialysis is a more scalable technology, as its use of porous media can enable high throughput without requiring the fabrication of many parallel microfluidic systems.\cite{deng2013} Another unique feature of shock ED is the ability to propagate the depletion zone controllably through micron-scale frit pores, enabling a tuneable ion depletion zone which can extend to millimeters or larger in length to further increase throughput. 

In this work, we demonstrate that our shock ED cell can perform a number of functions in addition to (and simultaneously with) water desalination, including filtration, disinfection, and ion separations (see Figure~\ref{fig:sketch}). Both filtration and disinfection are important processes in modern water purification plants.\cite{greenlee2009reverse,sauvet2007ashkelon} With our cell, we demonstrate the filtration of micron-scale particles and aggregates of nanoscale particles present in the feedwater, and we hypothesize this was due to steric hinderance by our microporous frit. We further demonstrate that approximately 99$\%$ of \textit{E. coli} bacteria placed in the feedwater were killed or removed upon flow through our shock ED prototype, illustrating the potential for in situ and additive-free disinfection in shock ED. In addition, we show that our prototype can continuously separate electrochemically inactive ions by charge, consistent with theory~\cite{rica2010} but contradicting claims that ICP acts as a ``virtual barrier" to all charged species~\cite{kim2010}.

\section{MATERIALS AND METHODS}
\subsection{Shock ED Device}
\begin{figure}
\includegraphics[width=5in]{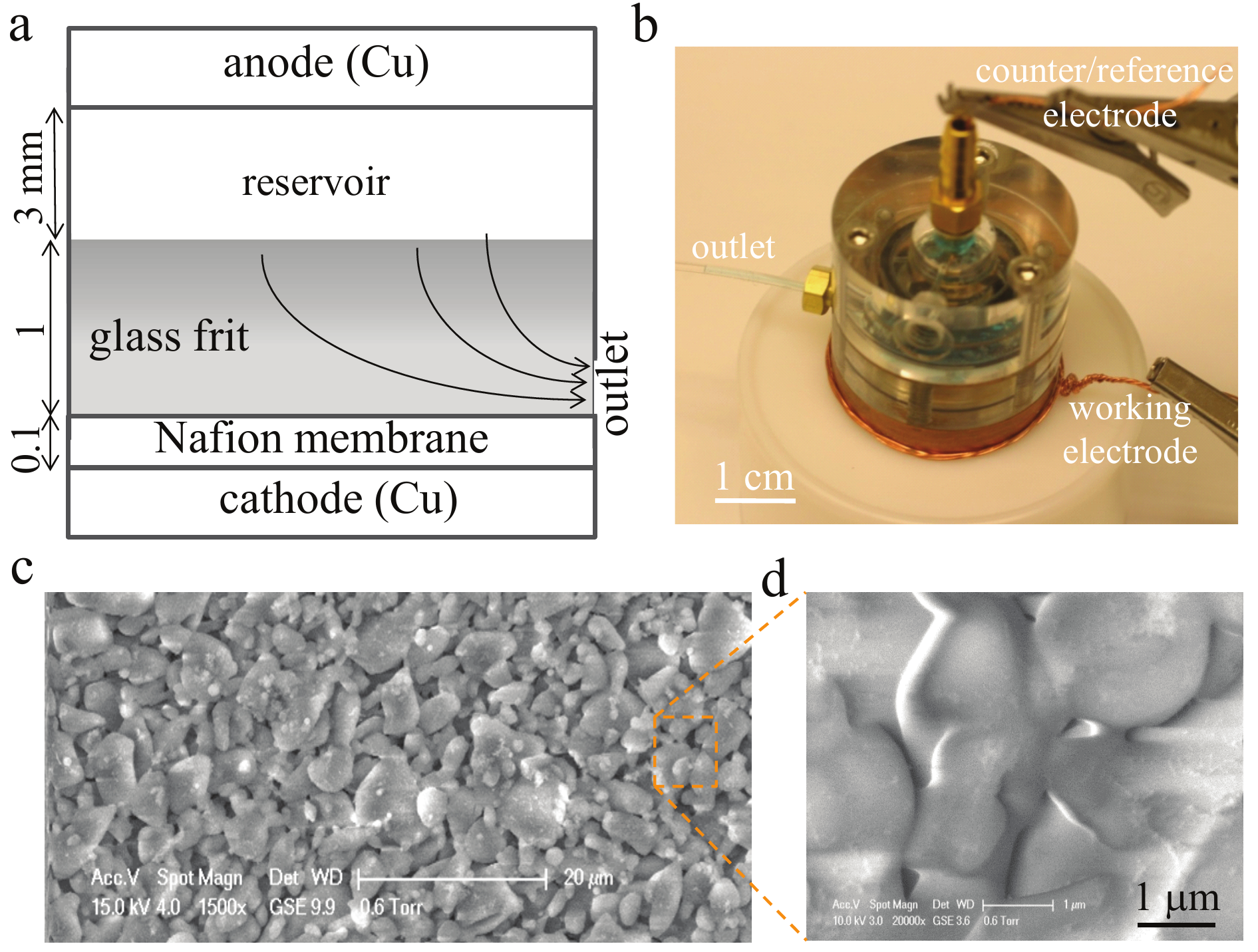}
\caption{Description of the prototype shock ED device used in this work. (a) A sketch of the frit/membrane/electrode sandwich structure (not to scale), where arrows indicate the fluid flow direction; (b) A photograph of shock ED device; (c) An SEM micrograph of glass frit showing its pore structure, and (d) enlarged micrographs indicating pore size around $500-700$ nm.}
\label{fig:device}
\end{figure}

A schematic and a photograph of our shock ED device are shown in Figures~\ref{fig:device}a and b. The setup consists of a cylindrical silica glass frit (Adams \& Chittenden Scientific Glass), which is 1~mm thick and has a 5~mm radius. The frit is placed against a Nafion membrane, and the membrane is in direct contact with the copper disk cathode. The frit is separated from the copper disk anode by a reservoir of copper sulfate (CuSO$_4$) electrolyte (3~mm thick reservoir). The frit has a random microstructure with pores roughly $500-700$ nm in diameter (Figure~\ref{fig:device}c, d), with an internal surface area (measured via BET) of $a_m=1.75$ m$^2$/g, and a density of $\rho_m=1.02$ g/cm$^3$. The pore surface charge is negative in these solutions and depends on the copper sulfate concentration, so that surface conduction and electro-osmotic flow promote the transport of positive copper ions to the cathode, leading to over-limiting current and deionization shocks~\cite{deng2013}. 

\subsection{Sample Preparation}
A 1~M CuSO$_4$ stock solution was prepared by dissolving $2.5$ grams of CuSO$_4$ (Science Company) into  $10$ ml of deionized water. This stock solution was further diluted $10$ times to obtain $0.1$ M CuSO$_4$ solution, and again diluted to obtain $1$ mM CuSO$_4$ solution. To demonstrate size-based filtration, two suspensions were prepared by adding 50~$\mu$m diameter green fluorescent polymer microspheres (Thermo Scientific) and 50~nm diameter red fluorescent nanoparticles (Thermo Scientific) into 1~mM CuSO$_4$ solution. The concentration of these suspensions were 20~mg/mL and 2~mg/mL, respectively. To demonstrate charged-based separation, positively and negatived charged fluorescent dye solutions were also prepared.  For positively charged dye, $2\times10^{-5}$~g/mL of Rhodamine B fluorescent dye was mixed into 1~mM CuSO$_4$. The pH of this solution was measured to be 4.2, far enough below Rhodamine's isoelectric point of 6 to ensure that the dye is positively charged~\cite{Schrum2000,Garcia2005b,Oh2008}. For negatively charged dye, 1~mg/mL of fluorescein dye (Sigma-Aldrich) was mixed into 1~mM CuSO$_4$~\cite{Buharova2008}. Isopropyl alcohol was added to ensure solubility. In order to evaluate the disinfection capabilities of the device, we prepared suspensions of \textit{Escherichia coli} K12 (ATCC). The bacteria were cultured in LB broth at 37$^\circ$ with shaking, were then resuspended in $1.5$ M NaCl solution during log phase. This concentration was selected to minimize osmotic shock upon transfer from the LB broth. After experiments were completed, the bacteria were stained with a BacLight live/dead staining kit as per the manufacturer's instructions (Invitrogen) and thus the live cells could be observed with a microscope.

\subsection{Device Operation}
An electrochemical analyzer (Uniscan instruments PG581) was used to apply voltage to the device. The analyzer's reference and counter electrode leads were connected to the anode, and the working electrode lead was connected to the cathode. After about $10$ minutes, the current reached steady state, and collection of fluid from the outlet of the device began. As indicated in Figure~\ref{fig:device}a, the fluid flow was directed towards the cathode side of the device (to force flow through the depletion zone) . Flow rate was precisely controlled by a syringe pump (Harvard Apparatus), and the fluid extraction time varied from several minutes to several tens of minutes in order to collect roughly 1~mL of fluid from the outlet for accurate post-experiment analysis. 

\subsection{Microscopy}
A Nikon Ti-U inverted fluorescence microscope, a 4X or 10X objective, and a Photometrics Coolsnap HQ2 CCD camera were used to capture optical micrographs of samples of fresh solutions or solutions from the outlet of the device. Illumination was provided by a mercury lamp (Intensilight) for the fluorescence images, and a standard lamp for the bright field images. To count live \textit{E. coli} cells or particles, samples were loaded into a hemacytometer (INCYTO). For \textit{E. coli}, the live/dead cell counting was done using a Nikon C-FL B-2E/C FITC Filter for the live bacteria (dyed with Syto BC, thus appearing green) and a Nikon C-FL Texas Red HYQ for the dead bacteria (dyed with Propidium Iodide, thus appearing red). Image analysis was performed using ImageJ software, either to count the number of particles, count live and dead cells, or to measure the integrated fluorescent intensity of the fluorescent dye solution. In experiments involving fluorescent dyes, any photobleaching effects were not significant as we confirmed that there was no decrease in the integrated fluorescence of a control sample kept next to the device over the course of the experiment.

\section{RESULTS AND DISCUSSION}
\subsection{Filtration Based on Size}
In contrast to electrodialysis,\cite{strathmann2004} shock electrodialysis utilizes a porous glass frit placed between ion exchange membranes to enable overlimiting current~\cite{dydek2011} and propagate a deionization shock wave~\cite{mani2009,mani2011,dydek2013,yaroshchuk2012acis}. The frit in a shock ED system can also be used as a filter to sterically exclude undesirable particles from the flow (for example, particulates in feedwater in a water purification process). The glass frit in our prototype has pore diameters between roughly 500 to 700 nm, so micron-scale particles can potentially be largely excluded from the fluid at the outlet.

\begin{figure}
\includegraphics[width=5in]{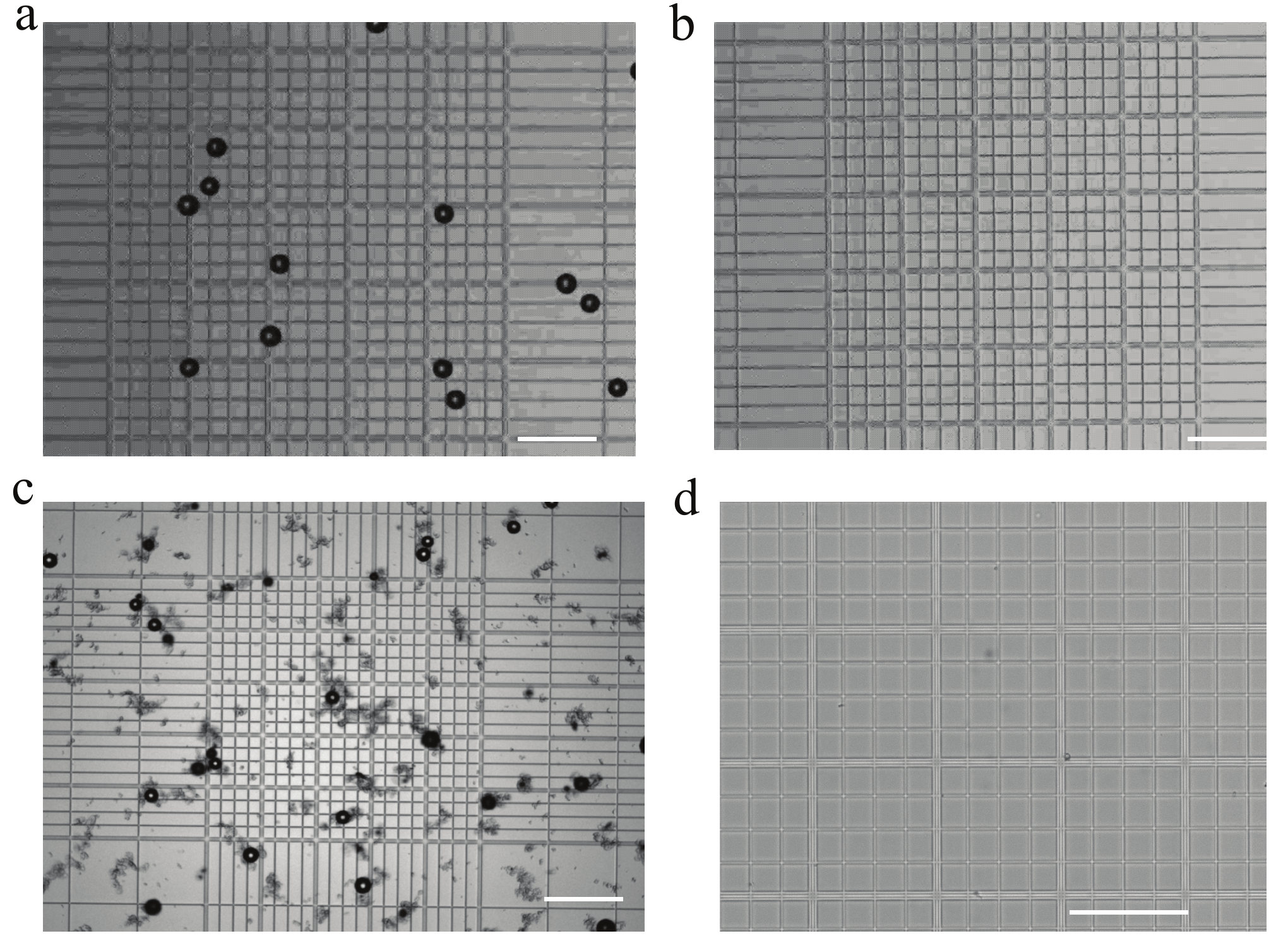}
\caption{Results demonstrating the size-based filtration of our shock ED device. Bright field images for the feedwater (a) and (c), and images on a sample from the corresponsing outflow (b) and (d). The particles with $50$ $\mu$m diameter were completely removed from inlet (a) to outlet (b). Aggregation of $50$-nm-diameter nanoparticles were also filtered from inlet (c) to outlet (d). Scale bars are $200$ $\mu$m.}
\label{fig:filtration}
\end{figure}

In the first experiment, the reservoir was filled with the suspension of 50~$\mu$m particles in copper sulfate solution. We then flowed this suspension through the shock ED device (from anode to cathode), without applying electric field, and captured the effluent. The microscopy image taken of the inlet solution is shown in figure \ref{fig:filtration}a, and here we can observe the particles as dark dots. As shown in figure \ref{fig:filtration}b, the absence of particles was observed in the outflow solution, demonstrating that our shock ED prototype successfully filtered out these particles. 

In a similar experiment, we investigated the behavior of particles with diameter less than the frit pore size using the suspension of $50$ nm-diameter particles in aqueous copper sulfate solution. From the image of the inlet solution in figure \ref{fig:filtration}c, the nanoparticles appeared to flocculate into aggregates due to surface interactions\cite{Hotze2010,Petosa2010}. Due to the spatial resolution limitations of optical microscopy, we could not observe directly any unaggregated $50$ nm-diameter particles potentially present in solution. However, we observed that the aggregates were large enough to be filtered by porous medium, as no aggregates were observed in the effluent samples (figure \ref{fig:filtration}d).  

\subsection{Disinfection}
Removing waterborne pathogens is a critical part of many water purification processes~\cite{shannon2008,schoen2010}. State-of-the-art reverse osmosis (RO) desalination plants utilize dedicated post-treatment procedures for water disinfection, typically the addition of chlorine or chlorine by-products, to eliminate most harmful microorganisms\cite{Huitle2008}. The shock ED device investigated here utilizes a glass frit with submicron-radius pores that can potentially serve as a filter to remove many biological species during deionization, eliminating the need for any post-treatments. In addition, we postulate that the high electric fields present in the deionization region may further reduce the survival rate of any microorganisms that make it through the glass frit.

\begin{figure}
\includegraphics[width=5in]{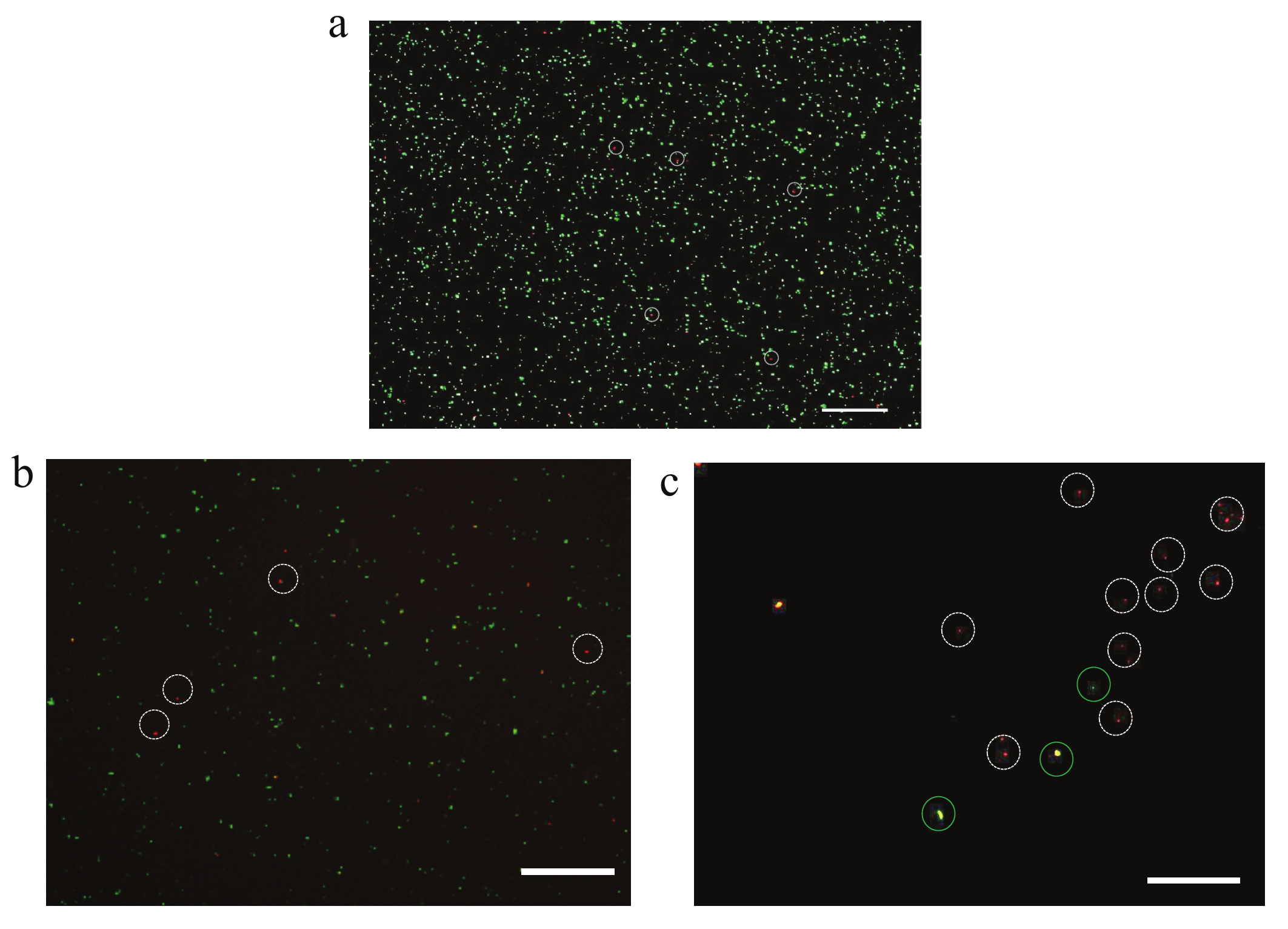}
\caption{Results demonstrating the disinfection of a feedwater containing \textit{E. Coli} as a model bacteria. The microscopy image of the feedwater (a), image of a sample from the outlet stream of the device without porous the porous frit (b), an outlet stream image from the device with porous frit included (c). The live bacteria fluoresce green, and the dead bacteria are red. Scale bar is $200, 50, 50 \mu m$, for (a)-(c) respectively.}
\label{fig:disinfection}
\end{figure}

To demonstrate the disinfection functionality in our device, we measured the change in concentration and viability of a suspension of \textit{E. coli} between the inlet and outlet of the device. For these experiments, to ensure bacteria viability, they were suspended in a $1.5$ M sodium chloride (NaCl) solution (see Materials and Methods section). A sample of the inlet \textit{E. coli} solution is shown in Figure~\ref{fig:disinfection}a. The bacteria were stained with a live-dead kit prior to microscopy, and so live bacteria appear green, and dead as red. Using our microscopy setup, we performed a cell counting analysis which showed that the inlet sample contains bacteria which were $99\pm0.4\%$ viable. Outlet samples were taken under two conditions: one with a flowrate of 50 $\mu$L/min through the device (and no voltage applied), shown in Figure~\ref{fig:disinfection}c, and one with the same flow rate but the glass frit removed, as shown in Figure~\ref{fig:disinfection}b. With the frit removed, cell viability was measured to be to $96.4\pm0.9\%$. With the frit in place, we measured a much reduced viability of the bacteria in the outlet sample, which was $28\pm8\%$. We also observed (with the frit in place) a strong reduction in the concentration of cells in the outlet sample, as shown in Figure~\ref{fig:disinfection}. Here, in the outlet sample, $96.5\pm1.8\%$ of bacteria were absent relative to the reservoir sample. Combined with the reduced viability of the outlet sample, roughly 1\% of initially viable bacteria remained present and viable in the outlet sample when the frit was in place. We hypothesize that the latter results are due to both steric exclusion of a majority of the bacteria from the frit pore space (\textit{E. coli} have typically micron-scale dimensions\cite{Asami1980}), and an inhospitable environment to the bacteria which are able enter the glass frit. 

We also hypothesize that the strong electric fields presumably created within the ion depletion zone can also reduce further the viability of the bacteria passing through the device.\cite{jeon2013} Previous works describe that an electric field of roughly $0.8\sim 2$ V/$\mu$m can promote cell death.\cite{Hulsheger1981},\cite{Grahl1996} However, we were unable to reliably test this hypothesis in this work, because the use of sodium chloride rather than copper sulfate likely inhibited the formation of a concentration shock (and thus a strong electric field) in our prototype. With sodium chloride (and copper electrodes), our device relied on water electrolysis electrode reactions rather than copper redox reactions to pass a current through the device, which can cause zones of perturbed pH to enter the system \cite{persat2009basic}. Future work will develop prototypes capable of generating shocks in sodium chloride and other general electrolytic solutions, allowing us to test the effect of depletion zone electric fields on bacteria survival.

\subsection{Separation Based on Charge}
In microfluidic devices leveraging ICP to perform molecular sample stacking~\cite{wang2005} and water desalination~\cite{kim2010}, the ion depletion zone was reported to act as a ``virtual barrier" for all charged particles (both positively and negatively charged). A possible mechanism for such charge-independent attraction is diffusio-phoresis~\cite{deryaguin1961,dukhin1974,prieve1984}, in which charged particles climb a salt concentration gradient in the absence of an applied electric field.  When current is being passed, however, this effect competes with classical  electrophoresis in the joint phenomenon of electro-diffusiophoresis studied by Malkin and A. Dukhin~\cite{malkin1982}.  Rica and Bazant have shown that electrophoresis generally dominates diffusiophoresis in large concentration gradients due to the enhanced electric field of the depleted solution~\cite{rica2010}, thus enabling separation by charge in particle flows through regions of ICP.    Recently, Jeon et al. analyzed the forces acting on charged particles flowing through the depletion region in microfluidic ICP with negatively charged channel walls and reached consistent conclusions, that negatively charged (counter-ionic) species are repelled from the depletion zone via strong electrophoretic forces, while positively charged species cannot be repelled.\cite{jeon2013} Jeon et al. further developed a device which separated negatively charged particles via the deflection of their path through the depletion zone (where the extent of deflection correlated to the particles' electrophoretic mobility).\cite{jeon2013} In this work, we show that ICP can in fact accelerate the passage of positively charged species, thus contradicting the virtual barrier claim and demonstrating, apparently for the first time, charge-based separation by the ion depletion zone

\begin{figure}
\includegraphics[width=5in]{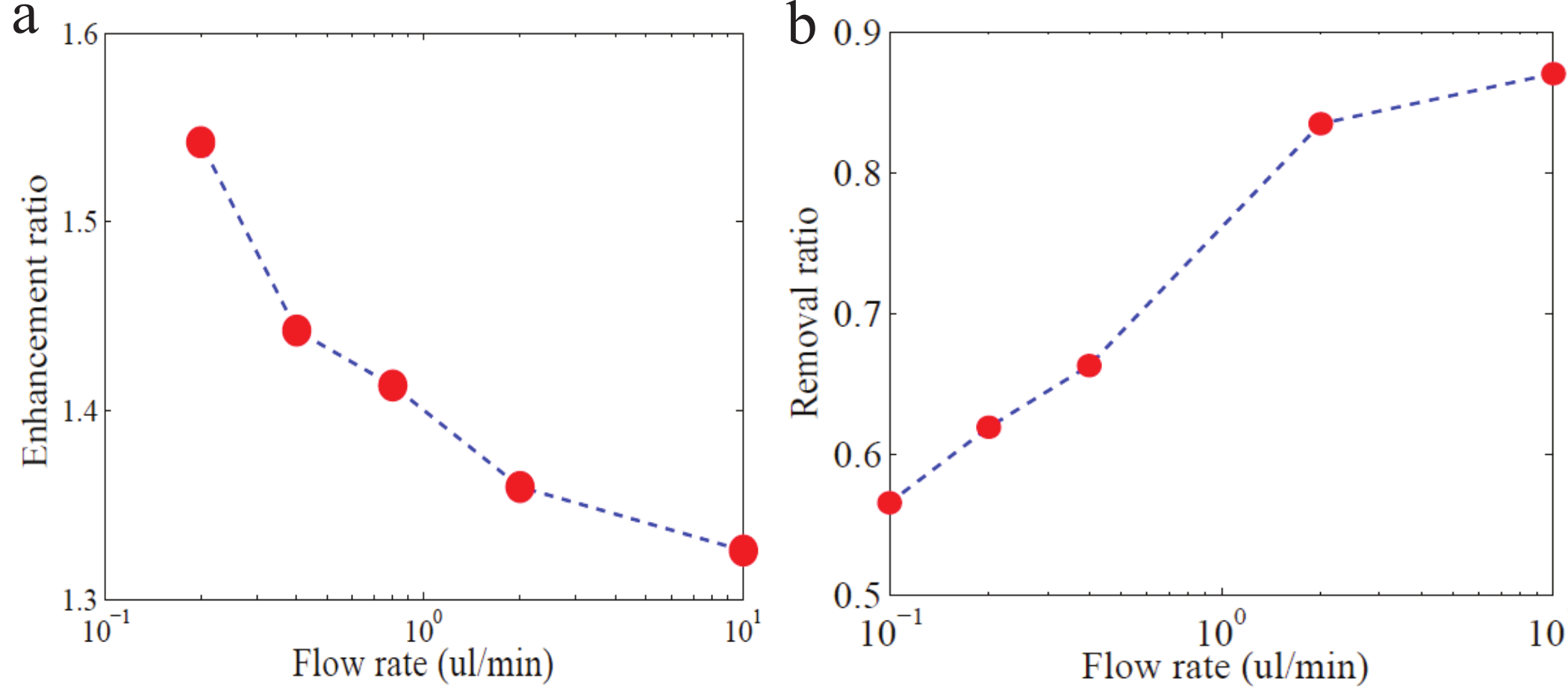}
\caption{Results demonstrating the charge-based separation of non-electrochemically active ions by our shock ED device. (a) The observed enhancement ratio of a fluorescent, positively charged dye versus flow rate of feewater through the device. Enhancement ratio is the ratio of dye concentration (fluorescent intensity) of the outlet sample to the feedwater sample. Enhancement ratio is always greater than unity, indicating that positively charged dye accumulated in the depletion zone of the device. Enhancement ratio decreases as flow rate is increased. (a) The observed removal ratio of a fluorescent, negatively charged dye versus flow rate. Removal ratio is also defined as the ratio of dye concentration (fluorescent intensity) of the outlet sample to the feedwater sample, but in this case, this ratio is always below unity indicating that negatively charged dye was repelled from the depletion zone. Removal ratio increases as flow rate is increased.}  
\label{fig:separation}
\end{figure}

In order to demonstrate effect of the depletion zone on positively charged species, we injected the positively charged Rhodamine dye solution into the device reservoir, and applied a constant voltage of 1.5~V. Note that Rhodamine is considered a non-electrochemically active species, as it does not participate in the electrode reactions (electrode reactions include the positively charged copper ions). Based on the applied electric field, the positively charged dye were expected to move via electrophoretic forces towards the cathode where they accumulate, forming an enrichment region near the outlet. The outlet and reservoir concentrations of Rhodamine were calculated by injecting samples into a hemacytometer chip, and measuring their integrated fluorescent intensities ($I$) using an optical microscope. As expected, the dye concentration at the outlet was observed to be greater than at that of the reservoir. Quantitatively, we defined the enhancement ratio of fluorescent intensity as $I_{outlet}/I_{inlet}$, where $I_{inlet}$ is the integrated fluorescent intensity of solution in the inlet or reservoir, and $I_{outlet}$ is the integrated fluorescent intensity of solution extracted from the outlet. This enhancement ratio is shown in Figure~\ref{fig:separation}a as a function of extraction flow rate. As the flow rate decreased, the effluent became more enriched in Rhodamine as the integrated fluorescent intensity of solution increased. When no voltage was applied to the shock ED device, as expected, no concentration enhancement was observed in the effluent solution.

The experiment was then repeated with the negatively charged fluorescein dye solution. In contrast to the positively charged dye, the fluorescein was expected to migrate electrophoretically towards the anode, thus being "repelled" from the depletion zone by the large electric field there. When the dye concentration was measured quantitatively at the inlet and outlet (at the cathode-side), we observed a depletion of fluorescein at the outlet as expected. The removal ratio, defined as $I_{inlet}/I_{outlet}$, is shown as a function of flow rate in Figure~\ref{fig:separation}b, and this ratio increased with increasing flow rate.

\subsection{Conclusion}
We have demonstrated that shock ED devices can perform many functions in addition to (and simultaneously along with) water desalination, including filtration, disinfection and separations by charge and by size.  As shock ED employs a microporous frit within the flow channel between the membranes, we were able to demonstrate steric filtration of microscale particles and aggregates of nanoscale particles. Further, we were able to kill or remove approximately $99\%$ of viable E Coli bacteria present in the feedwater upon flowing through the shock ED device with applied voltage. We hypothesize that further reductions in bacteria viability are achievable via the presumably strong electric fields present within the shock region, and future work will focus on building a prototype capable of testing this hypothesis. Lastly, we demonstrated that our shock ED device can continuously separate positively charged species from negatively charged ones that are small enough to pass into the porous frit. These demonstrated functionalities can be applied to water purification systems (filtration, disinfection), as well as particulate, molecular or biological separation systems, and demonstrate the potential of shock ED as a versatile new technique for chemical engineering. In order to avoid the need for Faradaic reactions, such as water splitting, to sustain the current, shock ED could also be performed with blocking porous electrodes to drive salt depletion, as in (membrane~\cite{biesheuvel2009MCDI}) capacitive deionization~\cite{porada2013,anderson2010}.

\section{AUTHOR INFORMATION}
\textbf{Corresponding Author}
Phone: 1-617-324-2036. E-mail:bazant@mit.edu.
\\*
\textbf{Notes}
The authors declare no competing financial interest.

\section{ACKNOWLEDGEMENTS}
This work was supported by grants from Weatherford International and the MIT Energy
Initiative. W. A. would like to thank the Department of Chemical Engineering of Ecole Polytechnique de Montreal for the internship support at MIT.
\newpage

\newpage

\newpage

\newpage

\newpage

\bibliography{elec42}

\end{document}